\documentclass[conference]{IEEEtran}
\usepackage{url}
\usepackage[utf8]{inputenc}
\usepackage{xcolor}
\usepackage{amsmath}
\usepackage{amssymb}
\usepackage{algorithm}
\usepackage{algorithmic}

\usepackage[acronyms,nonumberlist,nopostdot,nomain,nogroupskip]{glossaries}
\usepackage{tablefootnote}
\usepackage{booktabs}
\usepackage{tabularx}
\usepackage{tikz}
\usepackage{pgfplots}
\pgfplotsset{compat=newest}
\pgfplotsset{plot coordinates/math parser=false}
\usetikzlibrary{plotmarks,patterns,decorations.pathreplacing,backgrounds,calc,arrows,arrows.meta,spy,matrix,backgrounds}
\usepgfplotslibrary{patchplots,groupplots}
\usepackage{tikzscale}
\usepackage{enumitem}

\usepackage{multirow}
\usepackage[noadjust]{cite}

\usepackage{subfigure}

\usepackage[outdir=./img/]{epstopdf}
\graphicspath{{./img/}}
\DeclareGraphicsExtensions{.eps,.ps,.png}

\usepackage{mathtools}

\usepackage{dblfloatfix}    % To enable figures at the bottom of page
\usepackage{colortbl}

\DeclareUnicodeCharacter{2212}{-} % TEMPORARY, just for TikZ

\newcommand{\Hb}{\mathbf{H}}

\newcommand{\A}{\mathbf{A}}

\newcommand{\I}{\mathbf{I}}

\newcommand{\Phib}{\mathbf{\boldsymbol{\Phi}}}

\newcommand{\Xib}{\mathbf{\boldsymbol{\Xi}}}

\newcommand{\x}{\mathbf{x}}

\newcommand{\y}{\mathbf{y}}

\newcommand{\vphi}{\mathbf{\boldsymbol{\phi}}}

\newcommand{\rr}{\mathbf{r}}
\newcommand{\z}{\mathbf{z}}
\newcommand{\ab}{\mathbf{a}}

\newcommand{\Ex}[2]{{\textnormal{E}_{#1}\left[#2\right]}}

\usepackage{pifont}
\usepackage{theorem}

\theoremstyle{plain}

   \definecolor{blueH3}{rgb}{0,.5,1}
   \definecolor{blueH2}{rgb}{0,0.25,0.75}
   \definecolor{blueH1}{rgb}{0,0,0.5}   
   \definecolor{grayOldText}{rgb}{.5,.5,.5}
   \definecolor{VCobalt}{HTML}{005682}
   \definecolor{TZTeal}{HTML}{008080}
   \definecolor{KYJade}{HTML}{008151}
   \definecolor{ARust}{HTML}{a10000}
   \definecolor{FFucsia}{HTML}{7000c3}
   
\newcommand{\SWITCH}[1]{\STATE \textbf{switch} (#1)}
\newcommand{\ENDSWITCH}{\STATE \textbf{end switch}}
\newcommand{\CASE}[1]{\STATE \textbf{case} #1\textbf{:} \begin{ALC@g}}
\newcommand{\ENDCASE}{\end{ALC@g}}

\newcommand{\DEFAULT}{\STATE \textbf{default:} \begin{ALC@g}}
\newcommand{\ENDDEFAULT}{\end{ALC@g}}
\newcommand{\DEFAULTLINE}[1]{\STATE \textbf{default:} }
\begin{document}

\title{Compressed Sensing Channel Estimation for OTFS Modulation in Non-Integer Delay-Doppler Domain}

  \author{
  \IEEEauthorblockN{Felipe G\'omez-Cuba}
   
\IEEEauthorblockA{
    $^1$atlanTTic, University of Vigo, Spain. Email: \texttt{gomezcuba@gts.uvigo.es}\\
    }
}

\flushbottom
\setlength{\parskip}{0ex plus0.1ex}

\maketitle
\thispagestyle{empty}

\begin{abstract}
This paper introduces a Compressed Sensing (CS) estimation scheme for Orthogonal Time Frequency Space (OTFS) channels with sparse multipath. The OTFS waveform represents signals in a two dimensional Delay-Doppler (DD) orthonormal basis. The proposed model does not require the assumption that the delays are integer multiples of the sampling period. The analysis shows that non-integer delay and Doppler shifts in the channel cannot be accurately modelled by integer approximations. An Orthogonal Matching Pursuit with Binary-division Refinement (OMPBR) estimation algorithm is proposed. The proposed estimator finds the best channel approximation over a \textit{continuous} DD dictionary without integer approximations. This results in a significant reduction of the estimation normalized mean squared error with reasonable computational complexity.
\end{abstract}

\begin{IEEEkeywords}
OTFS, Compressed Sensing, Channel Estimation
\end{IEEEkeywords}

\section{Introduction}

Consumer demand for versatile wireless communication systems continues to rise. Release 16 of the Third Generation Partnership Project (3GPP) mobile wirelss standard has been completed recently \cite{3GPPNRoverall16}, and the 5G mobile standard continues to evolve quickly with new additions in each release. Orthogonal Time Frequency Space (OTFS) is a promising novel waveform proposed recently \cite{OTFSwcncprop,wei2020orthogonal}. The OTFS waveform is based on an explicit delay-Doppler (DD) representation of time-varying multipath channels. Unlike a fading Orthogonal Frequency Division Multiplexing (OFDM) channel, the  OTFS channel is quasi-static, i.e., the DD represtation of the multipath reflections does not change over long periods of time. Moreover, OTFS is in practice an extension of OFDM with the addition of an internal representation of Doppler (in the sense of a closed group). Thus the new modulation has great backward compatibility with current standards, and the potential to facilitate performance improvements such as Doppler robustness in high speed mobility systems or reducing the perceived channel variability with movement in multipath environments with sparse reflections.
%In addition, the fact that OTFS is closely related with an explicit representation of the physical reflections of the signal represents an opportunity to introduce improvements in channel estimation techniques, specially in \textit{sparse multipath channels} where the number of reflections is much lower than the number of transmitted symbols in a single frame.

The DD domain is a two-dimensional representation of time-varying signals based on the Discrete Zak Transform (DZT). This mathematical tool has been studied even prior to the proposition of OTFS modulation, for example in polyphase filtering \cite{564174}. Hadani \textit{et al} \cite{OTFSwcncprop} proposed an OTFS implementation based on the cascade of two transforms: first a 2D \textit{Symplectic Fourier Transform} (SFT) converts the DD domain into a time-frequency (TF) grid (similar to a Single-Carrier OFDM (SC-OFDM) scheme in 5G uplink modulations). After the 2D SFT, a second operation termed \textit{Hadamard Transform} (HT) converts the TF grid into the final time-domain modulated signal. This implementation is not unique, and the DD signal may be converted directly to the time domain using the Inverse DZT (IDZT) as shown in \cite{8516353,9399255}. A derivation of DD modulations from the fundamental principles of sampling is also given in \cite{9392379}. Several authors have analyzed OTFS transmitter and receiver design \cite{9369138,8647394,9181410}, channel estimation \cite{8671740,8647394,9181410,wei2021performance}, window design \cite{wei2021performance,8516353}, etc. In this paper, we highlight three aspects of OTFS channel estimation that have not received sufficient attention in the prior works:
\begin{enumerate}[label=\roman*)]
 \item The vast majority of prior literature has assumed the multipath delay can be approximated as an integer multiple of the sampling period \cite{9369138,8647394,9181410,8671740,wei2021performance,8516353}. We model the OTFS DD modulation on a cyclic-prefixed continuous time signal without assuming integer delays in the channel. We show that removing this assumption results in significant differences in the equivalent channel expression.
 \item Our model employs the direct IDZT synthesis method, rather than the more usual SFT-HT cascade method. We remark that our derivation is independent from the prior result in \cite{9399255}, but both agree in pointing out that non-integer delays modify the channel model significantly.
 \item Motivated by the observation that non-integer delays play an important role, we propose a Compressed Sensing (CS) channel estimation algorithm for OTFS systems with IDZT architecture that estimates the \textbf{continuous} physical DD multipath parameters. Although prior works have applied CS to OTFS channel estimation, the previous references have employed discrete delay dictionaries, resulting in substantially different models.
\end{enumerate}

CS studies the estimation of sparse signals from a limited number of observation samples \cite{Duarte2011}. By exploiting sparsity, it is possible to develop estimation schemes that outperform conventional techniques. For example, finer delay resolution than expected according to the Shannon-Nyquist sampling constraint is possible in sparse multipath channel estimation. CS has gained interest in recent years as it has enabled to exploit the very large number of antennas and bandwidth in Massive MIMO and mmWave architectures for 5G \cite{3GPPNRoverall16,8844996}. The DD representation of signals is sparse in both delay and Doppler dimensions, and the OTFS modulation is a candidate technology for beyond 5G systems. Therefore, the study of CS channel estimation for OTFS is a natural next step for channel estimation in 5G. Moreover, in this paper we show that CS can play a very important role in coping with the differences that arise in OTFS system models depending on whether or not an assumption is made that delays are integer multiples of the sampling period. For this we develop an extension to OTFS waveforms of the Orthogonal Matching Pursuit with Binary-division Refinement (OMPBR) algorithm first introduced in \cite{8844996}. Differently from most CS algorithms, OMPBR is not limited in resolution by a discrete dictionary, and can estimate with arbitrary resolution the true continuous values of delay and Doppler multipath parameters. Our results shows that this leads to significant improvement of the estimation error.

The rest of this paper is organized as follows: Section \ref{sec:system} describes the OTFS channel model. Section \ref{sec:estimation} describes the OMPBR estimation algorithm. 
%In Section \ref{sec:data} we recall typical receiver techniques based on channel estimation and their performance. 
Section \ref{sec:numeric} validates the results in simulation. And finally, Section \ref{sec:conclusions} concludes the paper.

\subsection{Notation}
Calligraphic letters denote sets. $|\mathcal{A}|$ denotes the cardinality of set $\mathcal{A}$. Bold uppercase and lowercase letters denote matrices and vectors, respectively. $\A^H$ is the Hermitian and $\A^{\dag}$ the Moore-Penrose pseudo-inverse $(\A^H\A)^{-1}\A^{H}$. $\|\A\|_n=\left(\sum_{i,j}|a_{i,j}|^n\right)^{\frac{1}{n}}$ is the $\ell_n$ norm of $\A$, and $\|\A\|=\|\A\|_2$.

\section{System Model}
\label{sec:system} 

We begin by considering a discrete-time signal $x[n]$ of length $L$ transmitted in continuous time using a pulse $p(t)$
\begin{equation}
\label{eq:xsinc}
x(t)=\sum_{n=0}^{L-1}x[n]p(t-nT_s),
\end{equation}
over a continuous time-variant channel with output satisfying
\begin{equation}
\label{eq:chanoutput}
y(t)=\mathcal{H}(x(t))=\sum_{i\in\mathcal{P}}a_i e^{-j2\pi\nu_i t} x(t-\tau_i)+r(t)
\end{equation}
where $r(t)\sim\mathcal{CN}(0,\sigma_z^2)$ is AWGN and $\mathcal{P}$ is a set of paths characterized by their complex gain $a_i\in\mathbb{C}$, Doppler shift $\nu_i\in[-\frac{V\Delta f}{2},\frac{V\Delta f}{2})$ and delay $\tau_i\in[0,DT_s)$. We assume $D$ and $V$ are integers selected to capture the maximum delay and Doppler of the channel, normalized with regard to the symbol period $T_s$ and inverse message duration $\Delta f =\frac{1}{LT_s}$. Without loss of generality, we assume the symbol length satisfies $L=DV$. In typical systems it is common to adopt easily divisible values of $L$ such as powers of $2$ \cite{3GPPNRoverall16}. Thus, the case $L>DV$ can be easily included in our model adopting suitable integers $D'>D$ and $V'>V$ satisfying $L=V'D'$.

In order to convert \eqref{eq:chanoutput} into a circular convolution we consider the \textit{periodic extension} of $x[n]$, denoted $x_p[n]=x[n \mod L]$. This results in the continous-time transmitted signal
\begin{equation}
\label{eq:xpextension}
\begin{split}
x_p(t)&=\sum_{n=0}^{L-1}x[n]\sum_{m=-\infty}^{\infty}p(t-nT_s-mLT_s)\\
&=\sum_{n=0}^{L-1}x[n]g_L(t/T_s-n).   
  \end{split}
\end{equation}
Since sinc interpolation and sampling can be regarded as dual operations when $T_s$ is adequately chosen, without loss of generality in this paper we assume $p(t)=\textnormal{sinc}(t/T_s)$, and thus $g_L(t)=e^{-j(L-1)t}\frac{\sin(2\pi L t)}{\sin(2\pi t)}$ is a \textit{Dirichlet} periodical pulse train \cite{564174,9392379,wei2021performance}. Other pulses and windows can be modeled in discrete time, applied on $x_p[n]$, as long as we interpret $T_s$ as the \textit{sampling} period, not a \textit{modulation symbol} period.

We define the DZT of $x[n]$ \cite{564174} as
$$Z_x[d,v]=\sum_{u=0}^{V-1}x[d+uD]e^{-j2\pi\frac{uv}{V}}.$$
noting that the DZT is superjective and $Z_x[d,v]=Z_{x_p}[d,v]$. In other words, the \textit{Inverse Discrete Zak Transform} (IDZT) \cite{564174} outputs the periodic extension of the discrete signal 
\begin{equation}
\label{eq:xpmod}
x_p[n]=\sum_{v=0}^{V-1}Z_x[n\mod D,v]e^{j2\pi\frac{\lfloor n/D\rfloor v}{V}}.
\end{equation}
However, since $x[n]$ is limited to the interval $\{0\dots L-1\}$ we can always unequivocally reconstruct $x[n]$ from $Z_x[d,v]$.

\begin{figure}[t]
 \centering \includegraphics[trim={1cm 2.5cm 1cm 2.5cm},width=.8\columnwidth]{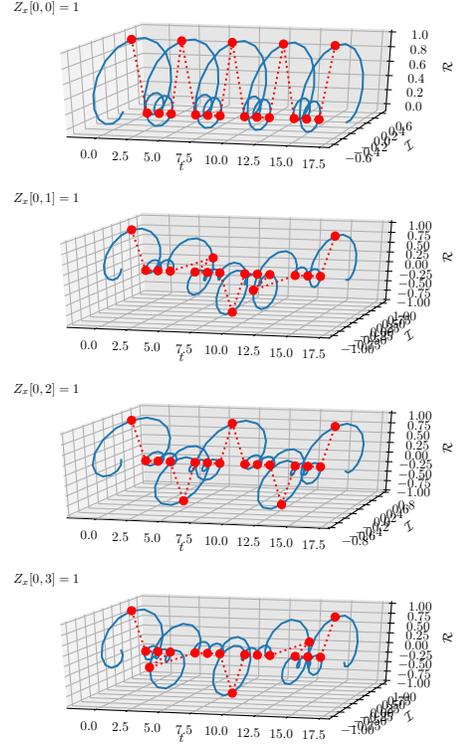}
 \caption{OTFS modulation orthonormal basis representation}
 \label{fig:pulse3D}
 \vspace{-.2in}
\end{figure}

To perform the OTFS modulation, information is encoded in the DD domain coefficients $Z_x[d,v]$ \cite{8516353}. The cyclic signal $x_p(t)$ is a linear combination of the OTFS coefficients, multiplied by an orthonormal basis formed by the Doppler shifted dirichlet pulses $e^{-j2\pi \frac{v}{V}t}g_D(t/T_s-d)$. Fig. \ref{fig:pulse3D} illustrates the orthonormal basis used in OTFS modulation \eqref{eq:xpmod}, for $D=V=4$ and $d=0$, with all Doppler shifts $v\in\{0\dots V-1\}$.

In practical implementations we would desire to synthesize a finite transmission. In order to replicate the circular convolution for a limited time, the transmitter sends the signal 
\begin{equation}
\label{eq:xcp}
x_{CP}(t)=\begin{cases}
           x_p(t)& t\in[-DT_s,LT_s]\\
           0&\textnormal{otherwise,}
          \end{cases}
\end{equation}
noting that $x_{CP}(t)$ is related to $x[n]$ by \eqref{eq:xpextension} and not \eqref{eq:xsinc}. Reciprocally, the receiver applies a rectangular window for the interval $t\in[0,LT_s]$ to \eqref{eq:chanoutput}, producing 
\begin{equation}
\label{eq:ycp}y_{CP}(t)=\begin{cases}
             \mathcal{H}(x_{p}(t))&\forall t\in[0,LT_s]\\
             0&\textnormal{otherwise.}\
            \end{cases}
\end{equation}

Finally, thanks to the cyclic prefix in $x_{CP}(t)$, when the receiver samples $y_{CP}(t)$ with period $T_s$ and computes its DZT, the result is an Equivalent Zak Channel (EZC) with output
\begin{equation}
 \label{eq:zakchan}Z_y[d,v]=\sum_{u'=0}^{V-1}\sum_{d'=0}^{D-1}h[d,v,d',v']Z_x[d',v']+Z_r[d,v]
\end{equation}
where $Z_r[d,v]$ is the Zak transform of the noise, which is still AWGN, and the channel gains are given by
\begin{equation}
 \label{eq:hall}
 \begin{split}
 h[d,v,d',v']=\sum_{i\in\mathcal{P}}&a_ie^{-j\nu_i d/D}g_D(d-d'-\tau_i/T_s)\\
 &\;\times g_V(v-v'-\nu_i/\Delta f)e^{-j2\pi \frac{v}{V} \lfloor \frac{d-d'}{D}\rfloor}.
 \end{split}
\end{equation}
Here we remark that the EZC output \eqref{eq:zakchan} cannot, in general, be expressed as a 2D convolution of $Z_h[d,v]$ and $Z_x[d,v]$, where $Z_h[d,v]$ is a DZT representation of the channel. There are two exceptions: when either all normalized Doppler coefficients $\nu_i/\Delta f$, or all normalized delay coefficients $\tau_i/T_s$, are integers. In either of these two special integer cases, we get
\begin{equation}
 \label{eq:hinteger}
 \begin{split}
 h[d,v,d',v']=Z_h[d-d',v-v'] 
 \end{split}
\end{equation}
where $Z_h[d,v]$ is the DZT of a Dirichlet channel probe
$$h[n]=\mathcal{H}(g_L(t/T_s))|_{t=nT_s}=\sum_{i\in\mathcal{P}}a_i e^{-j2\pi\nu_i nT_s} g_L(n-\tau_i/T_s).$$

We can define the vector $\x \in\mathbb{C}^L$ such that its $\ell$-th term contains $Z_x[\ell \mod D,\lfloor\ell/D \rfloor]$, and likewise $\y,\z\in\mathbb{C}^L$. Moreover, defining the matrix $\Hb \in\mathbb{C}^{L\times L}$ with $h[\ell \mod D,\lfloor\ell/D \rfloor,\ell' \mod D,\lfloor\ell'/D \rfloor]$ in its $\ell$-th row and $\ell'$-th column, the EZC \eqref{eq:zakchan} can be fully written as
\begin{equation}
 \label{eq:veczakchan}
 \y= \Hb\x+\z.
\end{equation}

Comparison of \eqref{eq:zakchan} and \eqref{eq:veczakchan} with a classic CP-OFDM system for time-invariant channels reveals some similarities and differences. In a CP-OFDM system with a $L$-DFT and an infinite cyclic extension of the transmitted signal, the Dirichlet pulse would also be observed in the discrete channel impulse response of length $D$ with non-integer multipath delays. Moreover, a second Dirichlet pulse would also arise as an interpolator in frequency, since the $L$ subcarrier gains would be correlated when $D<L$.

If we set $V=L$ and $D=1$ the OTFS scheme fully becomes CP-OFDM. More generally, for $D>1$ the IDZT/DZT transforms correspond to the time-domain half of the decimation-in-time FFT butterfly algorithm. This makes the implementation of the DZT computation very efficient. %In comparison, the first half of the decimation-in-frequency FFT algorithm can implement the Interleaved SC-OFDM waveform, and the 2D SFT proposed in \cite{OTFSwcncprop} can be interpreted as an OTFS to SC-OFDM conversion. 

In CP-OFDM channels, the cyclic extension enables writing the channel as a scalar multiplication in each subcarrier. However, due to non-integer Doppler, in the Zak channel we must take into account a grand total of $L^2$ multiplications to compute the first term of the EZC \eqref{eq:zakchan}. Owing to the highly structured expression \eqref{eq:hall}, a na\"ive and wasteful direct estimation of all $L^2$ elements of the matrix $\Hb$ would not be necessary. Instead, we assume that the number of multipath components is $P=|\mathcal{P}|$ satisfying $P\ll L$. Therefore, we can estimate the explicit multipath characteristics of the channel using CS methods as detailed in the next section. In a practical system, we would consider the finite implementation \eqref{eq:xcp} repeatedly transmitting consecutive blocks, each carrying $L$ symbols $\x$ on a physical signal of duration $(L+D)T_s$. Despite the time-domain channel being time-variant, the explicit multipath parameters are quasi-static. Thus, it suffices that the transmitter sends a known pilot $\x$ in the first block of each session.

\section{CS Channel Estimation Algorithm}
\label{sec:estimation} 

To design the CS estimator \cite{Duarte2011} we begin by defining the set of matrices $\Upsilon(\tau,\nu)$ for any value of $\tau$ and $\nu$, such that 
\begin{equation}
 \label{eq:chanbase}
 \begin{split}
    \{\Upsilon(\tau,\nu)\}_{d+vD,d'+v'D}=&e^{-j\nu_i d/D}g_D(d-d'-\tau/T_s)\\
    &\times g_V(v-v'-\nu/\Delta f)e^{-j2\pi \frac{v}{V} \lfloor \frac{d-d'}{D}\rfloor}.
 \end{split}
\end{equation}

We multiply this matrix by the known pilot $\x$ to obtain a generalized basis vector $\vphi(\tau,\nu)=\Upsilon(\tau,\nu)\x$.
We also define the 2D continuous channel spreading function as follows:
$$A_{\mathcal{P}}(\tau,\nu)=\sum_{i\in\mathcal{P}}a_i\delta(\tau-\tau_i,\nu-\nu_i).$$
So we may substitute \eqref{eq:chanbase} and \eqref{eq:hall} into \eqref{eq:veczakchan} \cite{9399255,wei2021performance}, producing 
\begin{equation}
 \label{eq:chansparse}
 \begin{split}
 \y&= \sum_{i\in\mathcal{P}}a_i\Upsilon(\tau_i,\nu_i)\x+\z\\ 
    &= \sum_{i\in\mathcal{P}}a_i\vphi(\tau_i,\nu_i)+\z\\ 
    &= \int_{0}^{DT_s}\int_{0}^{V\Delta f}\vphi(\tau,\nu)A_{\mathcal{P}}(\tau,\nu) d\nu d\tau+\z.\\
 \end{split}
\end{equation}
Where the interval $[\frac{V\Delta f}{2},V\Delta f)$ represents the negative Doppler ranges $[-\frac{V\Delta f}{2},0)$, due to the periodicity of $g_V()$. 

Since the integral in \eqref{eq:chansparse} may be challenging, it is frequent to design CS schemes using a discrete \textbf{dictionary} of values \cite{Duarte2011}. In our case, we define dictionaries of $K_\tau$ and $K_\nu$ evenly spaced points in the intervals $[0,DT_s)$ and $[0,V\Delta f)$, respectively. The dictionary is thus the cartesian product $\mathcal{D}_{\tau}\times \mathcal{D}_{\nu}$, where
\begin{equation}
\label{eq:Dtaudef}
  \mathcal{D}_{\tau}=\{0\dots \frac{K_{\tau}-1}{K_{\tau}}DT_s\},
\end{equation}
and
\begin{equation}
\label{eq:Dnudef}
    \mathcal{D}_{\nu}=\{0\dots \frac{K_{\nu}-1}{K_{\nu}}V\Delta f\}.
\end{equation}

Finally, we define the dictionary basis matrix containing all possible basis vectors in the dictionary as
$$\Phib_{\mathcal{D}_{\tau}\times \mathcal{D}_{\nu}}=(\vphi(0,0),\dots,\vphi(\frac{K_{\tau}-1}{K_{\tau}}DT_s,\frac{K_{\nu}-1}{K_{\nu}}V\Delta f)),$$
and an approximation of the channel output can be written as
\begin{equation}
 \label{eq:chandiscretesparse}
 \begin{split}
    \y\simeq& \Phib_{\mathcal{D}_{\tau}\times \mathcal{D}_{\nu}}\ab+\z
 \end{split}
\end{equation}
where $\ab$ is a sparse vector with non-zero values in the indices that correspond to existing paths in the channel.

We remark that the dictionary dimensions $K_{\tau}$ and $K_{\nu}$ may be greater than $D$ and $V$, respectively. In CS, this is referred to as an \textit{overcomplete} dictionary. Overcomplete dictionaries enable \textit{superresolution}, i.e., to resolve the support parameters, in our case delay and Doppler, with more resolution than it would be possible in non-sparse estimation according to the Shannon-Nyquist sampling constraints  \cite{Duarte2011}. In fact, we have noted that the channel expression \eqref{eq:zakchan} cannot be expressed in the form \eqref{eq:hinteger} when delay and Doppler are both fractional. Thus the use of orthogonal dictionaries without superresolution, with  $K_{\tau}=D$ and $K_{\nu}=V$, will lead to significant errors.

When we assume $K_{\tau}>D$ and $K_{\nu}>V$, the matrix $\Phib_{\mathcal{D}_{\tau}\times \mathcal{D}_{\nu}}$ in \eqref{eq:chandiscretesparse} is wide and does not have a pseudoinverse. Thus we cannot resort to well known methods such as Least Squares (LS) to estimate $\ab$. Therefore, the sparsity of $\ab$ must be exploited. The fundamental form of a CS estimator of $\ab$ is
\begin{equation}
\label{eq:cscanonical}
\hat\ab=\arg\min \|\ab\|_0 s.t. \|\y-\Phib_{\mathcal{D}_{\tau}\times \mathcal{D}_{\nu}}\ab\|^2_2\leq \xi 
\end{equation}
where $\xi$ is a tuning parameter to account for the noise. As problem \eqref{eq:cscanonical} is combinatorial, there are two general families of CS algorithms to convert \eqref{eq:cscanonical} into a tractable problem: greedy approximations of the combinatorial, or problem relaxations by substituting the $\ell_0$ norm with an $\ell_1$ norm \cite{Duarte2011}. A second distinction can be made between algorithms that assume the number of non-zero elements of $\ab$, $P$, is known and those that do not. Among greedy algorithms with unknown $P$, Orthogonal Matching Pursuit (OMP) variants are well known. Greedy algorithms with known $P$ include Compressive Sampling Matching Pursuit (CoSaMP). $\ell_1$ algorithms with unknown $P$ include Basis Pursuit De Noising (BPDN) and the Dantzig-Selector. Finally, the Least Absolute Shrinkage and Selection Operator (LASSO) solves a dual problem with known $P$, minimizing the $\ell_2$ norm with an $\ell_1$ constraint.

\begin{algorithm}[b]
 \caption{Orthogonal Matching Pursuit}
 \label{alg:omp}
 \begin{algorithmic}[1]
    \STATE Initialization $\hat{\mathcal{P}}=\emptyset$, $\rr=\y$
    \WHILE{$\|\rr\|^2>\xi$}
        \STATE $\displaystyle (\hat{\tau},\hat{\nu})=\underset{\tau\in\mathcal{D}_{\tau},\nu\in\mathcal{D}_{\nu}}{\arg\max}\|\vphi(\tau,\nu)^H\rr\|$
        \STATE $\hat{\mathcal{P}}=(\hat{\tau},\hat{\nu})\bigcup \hat{\mathcal{P}}$
        \STATE $\hat{\ab}=\Phib_{\hat{\mathcal{P}}}^\dag\y$
        \STATE $\rr=\y-\Phib_{\hat{\mathcal{P}}}\hat{\ab}$
    \ENDWHILE
    \STATE \textbf{Output} $\hat{\mathcal{P}}$, $\hat{\ab}$
 \end{algorithmic}
\end{algorithm}

Since the number of multipath reflections can change, in this paper we assume $P$ is not known. Moreover, to develop our extension using the continuous sparse model \eqref{eq:chansparse}, we must adopt a greedy algoritm. Therefore, of all algorithms in literature, the most relevant baseline for our study is OMP (Alg. \ref{alg:omp}). Here, we use the notation $\hat{\mathcal{P}}$ to represent the estimated set of delay and Doppler support values, which is increased by exactly one element in each iteration. Moreover, we use the matrix notation $\Phib_{\hat{\mathcal{P}}}$ to represent the support matrix associated with $\hat{\mathcal{P}}$, that is, each column of $\Phib_{\hat{\mathcal{P}}}$ is the vector $\vphi(\hat{\tau}_j,\hat{\nu}_j)$ associated with the $j$-th element of $\hat{\mathcal{P}}$. As the greedy algorithm increases the support, it builds $\Phib_{\hat{\mathcal{P}}}$ as a tall matrix containing a subset of the columns of $\Phib$, such that the LS operation in line 5 is possible. In each iteration, the residual $\rr$ contains the part of $\y$ that is orthogonal to the current basis $\Phib_{\hat{\mathcal{P}}}$ (line 6). Finally, the greedy algorithm adds an element to $\hat{\mathcal{P}}$ in each iteration (line 4), using a maximum residual correlation criterion (line 3), until the distance constraint $\xi$ is satisfied (line 2). The number of iterations equals the estimated number of paths $|\hat{\mathcal{P}}|$.

To choose the tuning parameter $\xi$ we study the Normalized Mean Squared Error (NMSE) expressed as follows:
 \begin{align}
NMSE&=\Ex{\Hb}{\frac{\|\Hb-\sum_{j\in\hat{\mathcal{P}}}\hat{a}_j\Upsilon(\hat{\tau}_j,\hat{\nu}_j)\|^2}{\|\Hb\|^2}}
 \label{eq:mse}\\
   &\simeq\Ex{\Hb}{\frac{\|(\y-\Phib_{\hat{\mathcal{P}}}\hat{\ab}\|^2}{\|\Hb\x\|^2}}
 \label{eq:mse2}\\
   &=\Ex{\Hb}{\frac{\|(\I-\Phib_{\hat{\mathcal{P}}}\Phib_{\hat{\mathcal{P}}}^\dag)(\Phib_{\mathcal{P}}\ab+\z)\|^2}{\|\Hb\x\|^2}}\label{eq:mse3}\\
   &\geq\Ex{\Hb}{\frac{\|\Xib_{\hat{\mathcal{P}}}\Phib_{\mathcal{P}}\ab\|^2}{\|\Hb\x\|^2}}+|\hat{P}|\sigma^2\Ex{\Hb}{\|\Hb\x\|^{-2}}\label{eq:mse4}
 \end{align}
where \eqref{eq:mse2} assumes that the average energy of the matrix $\Hb$ is symmetrically distributed in all directions, and we define $\Xib_{\hat{\mathcal{P}}}=\I-\Phib_{\hat{\mathcal{P}}}\Phib_{\hat{\mathcal{P}}}^\dag$. From the lower bound \eqref{eq:mse4} we note that the error contains two terms. The first term represents the projection of the true channel orthogonal to the basis $\Phib_{\hat{\mathcal{P}}}$, and decreases in each iteration. The second term represents the projection of the noise over $\Phib_{\hat{\mathcal{P}}}$, and grows in each iteration. Therefore, a natural choice for $\xi$ is making the algorithm stop when ``continuing more iterations would capture more noise than channel'', which results in $\|\rr\|^2<L\sigma^2\triangleq \xi$ \cite{8844996}.

The classic OMP algorithm displays three shortcomings: 
\begin{enumerate}[label=\roman*)]
 \item It is based on a discrete approximation \eqref{eq:chandiscretesparse} of a continuous truth \eqref{eq:chansparse}. We have argued that using \eqref{eq:hall} in \eqref{eq:zakchan} is not the same as using \eqref{eq:hinteger}. This mean that, in OTFS, quantized multipath integers can severely misrepresent the channel.
 \item Even though the dictionary sizes $K_\tau$ and $K_\nu$ can be increased to mitigate problem \textit{i)}, this results in a linear growth of the complexity of Line 3, which is equivalent to $\Phib_{\mathcal{D}_{\tau}\times \mathcal{D}_{\nu}}^H\rr$, that is, $K_\tau K_\nu L$ products.
 \item Whenever $\x$ changes the matrix $\Phib_{\mathcal{D}_{\tau}\times \mathcal{D}_{\nu}}$ needs to be precomputed. Performing $K_\tau K_\nu$ times the operation $\vphi(\tau,\nu)=\Upsilon(\tau,\nu)\x$ results in $K_\tau K_\nu L^2$ products.
\end{enumerate}

Due to the above, we introduce the OMPBR algoritm (Alg. \ref{alg:ompbr}). The main goal of OMPBR is to reduce the dictionary size in line 3 of Alg. \ref{alg:omp}. For this we observe that the correlation $f(\tau,\nu)=\|\vphi(\tau,\nu)^H\rr\|$ is not concave in the full search space $[0,DT_s)\times[0,V\Delta f)$, but it can be assumed to be locally symmetric in small regions near the maximum. Therefore, we perform a two step search of the correlation maximum: first, we search for the best ``local bin'' of size $T_s\times \Delta f$ in line 3. This search uses an auxiliar dictionary with $D\times V$ integer elements, which can be accelerated using \eqref{eq:hinteger}. Second, we identify the best small delay and Doppler offsets contained in the ``local bin'', in line 4. This search is implemented using the Binary-division Refinement Alg. \ref{alg:binref}. When $f(\tau,\nu)$ is symmetric around a local maximum, BR guarantees a distance $2^{-N_{ref}}$ to the true maximum in $N_{ref}$ iterations \cite{8844996}.

\begin{algorithm}
 \caption{OMP with Binary division Refinement (OMPBR)}
 \label{alg:ompbr}
 \begin{algorithmic}[1]
    \STATE Initialization $\hat{\mathcal{P}}=\emptyset$, $\rr=\y$
    \WHILE{$\|\rr\|>\xi$}
        \STATE $\displaystyle (\hat{d},\hat{v})=\underset{\substack{d\in\{0\dots D-1\}\\v\in\{0\dots V-1\}}}{\arg\max}\|\vphi(dT_s,v\Delta f)^H\rr\|$
        \STATE $\small\displaystyle (\hat{\mu}_{\tau},\hat{\mu}_{\nu})=\underset{\mu_\tau,\mu_\tau\in[\frac{-1}{2},\frac{1}{2}]}{\arg\max}\|\vphi((\hat{d}+\mu_\tau)T_s,(\hat{v}+\mu_\nu)\Delta f)^H\rr\|$
        \STATE $\hat{\mathcal{P}}=((\hat{d}+\hat{\mu}_{\tau})T_s,(\hat{v}+\hat{\mu}_{\nu})\Delta f)\bigcup \hat{\mathcal{P}}$
        \STATE $\hat{\ab}=\Phib_{\hat{\mathcal{P}}}^\dag\y$
        \STATE $\rr=\y-\Phib_{\hat{\mathcal{P}}}\hat{\ab}$
    \ENDWHILE
    \STATE \textbf{Output} $\hat{\mathcal{T}}$, $\hat{\mathcal{V}}$, $\hat{\ab}$
 \end{algorithmic}
\end{algorithm}

BR can also be employed to refine OMP decisions for an overcomplete auxiliar dictionary, although this drops the possibility of accelerating line 3 for integer dictionaries. OMPBR can be regarded as an approximation heuristic to classic OMP with an extremely large dictionary. By observing \eqref{eq:Dtaudef} and \eqref{eq:Dnudef}, if we define the superresolution factors $\kappa_\tau=\frac{2^{N_{ref}}K_{\tau}}{D}$ and $\kappa_\nu=\frac{2^{N_{ref}}K_{\nu}}{V}$, it is clear that the complexity of OMPBR scales much better with resolution. OMP ($N_{ref}=0$) finds the global correlation maximums with linear scaling in resolution $\Theta(\kappa_\tau\kappa_\nu)$, whereas OMPBR provides a good approximation of correlation decisions with only logarithmic scaling $\Theta(\log(\kappa_\tau\kappa_\nu))$.

\begin{algorithm}
 \caption{Binary division Refinement (BR)}
 \label{alg:binref}
 \begin{algorithmic}[1]
    \STATE \textbf{Input:} 2D function $f(x,y)$, initial point $(x_o,y_o)$
    \STATE Initialize $x_{up}=x_o+0.5$, $x_{down}=x_o-0.5$,
    \STATE $y_{up}=y_o+0.5$, and $y_{down}=y_o-0.5$
    \FOR{$N_{ref}$ iterations}
        \STATE $x_{mid}=(x_{up}+x_{down})/2$, $y_{mid}=(y_{up}+y_{down})/2$
        \SWITCH{Select maximum and halve search region}
        \CASE{$f(x_{up},y_{up})$ is maximum}
            \STATE $x_{down}=x_{mid}$, $y_{down}=y_{mid}$
        \ENDCASE
        \CASE{$f(x_{down},y_{up})$ is maximum}
            \STATE $x_{up}=x_{mid}$, $y_{down}=y_{mid}$
        \ENDCASE
        \CASE{$f(x_{up},y_{down})$ is maximum}
            \STATE $x_{down}=x_{mid}$, $y_{up}=y_{mid}$
        \ENDCASE
        \CASE{$f(x_{down},y_{down})$ is maximum}
            \STATE $x_{up}=x_{mid}$, $y_{up}=y_{mid}$
        \ENDCASE
        \ENDSWITCH
    \ENDFOR
    \STATE \textbf{Output} $(x_{mid},y_{mid})$
 \end{algorithmic}
\end{algorithm}

% \section{Data Transmission}
% \label{sec:data}
% 
% In the block channel model \eqref{eq:veczakchan} with block period $(D+L)T_s$, we assume a known pilot $\x_p$ is transmitted on the first block, producing the channel estimation $\hat{\Hb}=\sum_{i\in\hat{\mathcal{P}}}\hat{a}_i\Upsilon(\hat{\tau}_i,\hat{\nu}_i)$. Afterwards, in the second block and further, a data modulation $\x_{data}$ can be sent in the same channel \eqref{eq:veczakchan}. Since the channel estimation is imperfect, we write this as
% $$\y=\hat{\Hb}\x+(\Hb-\hat{\Hb})\x_{data}+\z.$$
% Here, we can lower bound receiver performance by modeling an AWGN channel where the channel matrix is truly $\hat{\Hb}$ and the AWGN is i.i.d. with variance $\sigma^2_{eq}=\frac{\Ex{}{\|^2(\Hb-\hat{\Hb})\x_{data}}\|^2}{L}+\sigma^2$.
% 
% In essence, the known channel creates known inter-symbol interference (ISI), whereas channel estimation error creates unknown ISI, and the receiver adopts a Treating Interference as Noise (TIN) scheme for the unknown part only. Under this conservative scheme, the Achievable Rate (AR) equivalent mutual information is well known to be of the form
% 
% $$AR=\frac{1}{L}\log_2(1+\frac{P(1-NMSE)}{\sigma^2_Z(1+NMSE)}).$$
% 
% Moreover, if we assume the common linear MMSE equalizer, producing
% $\s=(\hat{\Hb}^H\hat{\Hb}+\sigma_{eq}^2)^{-1}\hat{\Hb}^H\y$
% followed by a symbol-by-symbol decisor, we can compute the symbol error rate analytically. For example, for a QPSK it would be
% $$SER_{QPSK}=\mathcal{Q}(\sqrt{\frac{d_{mean}}{2}})$$
% where $d_{mean}=\Ex{\Hb}{\|\s-\x\|^2}$

\section{Simulations}
\label{sec:numeric} 

We simulate an OTFS system with dimensions D=V=16, modelled as in \eqref{eq:zakchan}-\eqref{eq:hall}. We generate 1000 random multipath channels with $P=3$ independent reflections with normalized distributions $\frac{\tau_i}{T_s}\sim U(0,D)$, $\frac{\nu_i}{\Delta f}\sim U(0,V)$, $\overline{a}_i\sim\mathcal{CN}(0,1)$, and finally normalize the gains as $a_i=\frac{\overline{a}_i}{\sqrt{\sum_{i\in\mathcal{P}}|\overline{a}_i|^2}}$. The pilot signals are random i.i.d. Gaussian sequences $\x\sim\mathcal{CN}(0,\I_L)$.

\begin{figure}[t]
 \centering \includegraphics[trim={.5cm 1cm .5cm 1cm},width=.75\columnwidth]{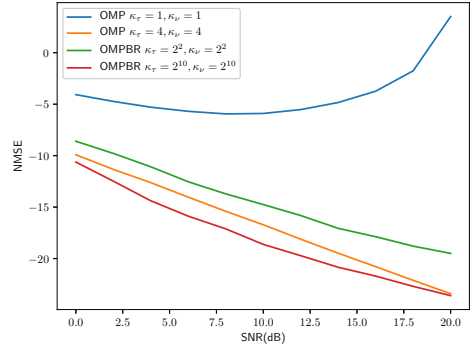}
 \caption{Channel NMSE}
 \label{fig:hmse}
 \vspace{-.05in}
\end{figure}

\begin{figure}[t]
 \centering
 \includegraphics[trim={.5cm 1cm .5cm 1cm},width=.75\columnwidth]{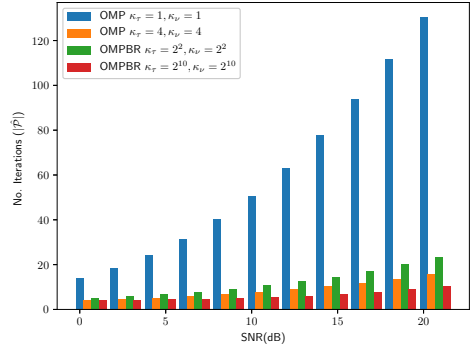}
 \caption{Number of estimated paths $|\hat{\mathcal{P}}|$}
 \label{fig:Npath}
 \vspace{-.2in}
\end{figure}

Fig. \ref{fig:hmse} represents the average NMSE evaluated as \eqref{eq:mse} without approximations, and Fig. \ref{fig:Npath} represents the number of estimated coefficients. One key observation is that OMP with $K_\tau=D$ and $K_\nu=V$ performs very poorly. In fact, its NMSE does not even decrease with SNR. The explanation is that, as we noted, expressions \eqref{eq:hall} and \eqref{eq:hinteger} are theoretically different. The greedy algorithm runs until it reaches the maximum number of iterations without ever converging and, since each iteration captures some noise, this results in large NMSE. Next, we consider OMP with $K_\tau=4D$ and $K_\nu=4V$. The NMSE of this estimator is significantly improved, and the number of estimated paths is reasonable. A slight overestimation of the number of paths is acceptable as discrete dictionaries can only approximate \eqref{eq:hall} so much. However, the use of a very large dictionary size can be computationally demanding. Next, we consider the case of OMPBR with $K_\tau=D$ and $K_\nu=V$ and $N_{ref}=2$. This estimator has the same search space and resolution as OMP with $\kappa_\tau=\kappa_\nu=4$, but stores a much smaller dictionary in memory. Remarkably, even though an orthogonal auxiliar dictionary is used, OMPBR performs quite well and does not fail catastrophically as OMP with the orthogonal dictionary. Since the BR search is local its NMSE is a bit worse than OMP with the same superresolution factors $\kappa_\tau=\kappa_\nu=4$. The number of estimated paths is slightly higher than the overcomplete OMP case as well. Finally, OMPBR with $K_\tau=D$ and $K_\nu=V$ and $N_{ref}=10$ has a superresolution factor of $\kappa_\tau=\kappa_\nu=1024$, which would be impractical with OMP, as it would require $10^6$ times more run time and memory. With an increase in complexity that only scales logarithmically, the 2D search space is increased by six orders of magnitude, and the NMSE is significantly improved compared with either OMPBR or OMP with coarser resolution. The number of necessary iterations ($|\hat{\mathcal{P}}|$) is also significantly lower than OMP with $\kappa_\tau=\kappa_\nu=4$.

% \begin{table*}
%  \centering
%  \caption{Total simulation runtime}
%  \label{tab:runtime}
%  \begin{tabular}{ccccc}
%     & OMP $\kappa_\tau=\kappa_\nu=1$& OMP $\kappa_\tau=\kappa_\nu=4$& OMPBR $\kappa_\tau=\kappa_\nu=4$& OMPBR $\kappa_\tau=\kappa_\nu=1024$\\
%   With computation of $\Phib_{\mathcal{D}_{\tau}\times \mathcal{D}_{\nu}}$  & OMP $\kappa_\tau=\kappa_\nu=1$& OMP $\kappa_\tau=\kappa_\nu=4$& OMPBR $\kappa_\tau=\kappa_\nu=4$& OMPBR $\kappa_\tau=\kappa_\nu=1024$\\
%   With precomputed $\Phib_{\mathcal{D}_{\tau}\times \mathcal{D}_{\nu}}$  & OMP $\kappa_\tau=\kappa_\nu=1$& OMP $\kappa_\tau=\kappa_\nu=4$& OMPBR $\kappa_\tau=\kappa_\nu=4$& OMPBR $\kappa_\tau=\kappa_\nu=1024$\\
%  \end{tabular}
% \end{table*}

\begin{figure}[t]
 \centering
 \includegraphics[trim={.5cm 1cm .5cm 1cm},width=.75\columnwidth]{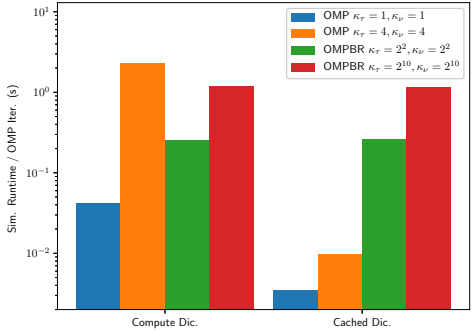}
 \caption{Total simulation runtime}
 \label{fig:runtime}
 \vspace{-.2in}
\end{figure}

To discuss the computational complexity in detail, we depict the mean simulation runtime per iteration in Fig. \ref{fig:runtime} (using a Dell XPS 13 2019 laptop). Our simulator implements a cache system to avoid precomputing the same matrix $\Phib_{\mathcal{D}_{\tau}\times \mathcal{D}_{\nu}}$ twice. We first focus on the points when the cache was not employed ($8\%$ of the total). The time to generate the matrix $\Phib_{\mathcal{D}_{\tau}\times \mathcal{D}_{\nu}}$ and load it in memory is counted in the total run time of Algs. \ref{alg:omp} and \ref{alg:ompbr} and divided by $|\hat{\mathcal{P}}|$. In this case, the run time of OMP with $\kappa_\tau=\kappa_\nu=1$ versus $\kappa_\tau=\kappa_\nu=4$ grew by a factor of $16$, as we have predicted. Moreover, the run-time of OMP with $\kappa_\tau=\kappa_\nu=4$ was an order of magnitude greater than OMPBR for the same resolution. Finally, the run time of OMPBR with $\kappa_\tau=\kappa_\nu=1024$ was $2.5$ times higher than OMPBR with $\kappa_\tau=\kappa_\nu=4$, confirming our analysis. Remarkably, OMPBR with $\kappa_\tau=\kappa_\nu=1024$ is faster than OMP with $\kappa_\tau=\kappa_\nu=4$, in addition to achieving better channel estimations.

When the matrix $\Phib_{\mathcal{D}_{\tau}\times \mathcal{D}_{\nu}}$ is cached, the run-time comparison changes significantly. The run time of classic OMP decreases by two orders of magnitude, whereas BR does not benefit from caching. Therefore, for moderate values of $\kappa_\tau$ and $\kappa_\nu$, for the subset of systems with unchanging pilot patterns and a large memory, classic OMP is still preferrable over OMPBR in terms of complexity. Still, OMP with a superresolution factor of $\kappa_\tau=\kappa_\nu=1024$ would be impossible to implement. OMPBR displays the better scaling and is the only estimator that supports extremely large values of $\kappa_\tau$ and $\kappa_\nu$, approximating the continuous truth of the channel \eqref{eq:chansparse}.

% \begin{figure}
%  \centering
%  \includegraphics[width=.85\columnwidth]{QPSKSER}
%  \caption{Channel NMSE}
% \end{figure}
% 
% \begin{figure}
%  \centering
%  \includegraphics[width=.85\columnwidth]{Rate}
%  \caption{Channel NMSE}
% \end{figure}

\section{Conclusions and Future Work}
\label{sec:conclusions} 

The OTFS waveform allows to represent signals in a sparse Delay-Doppler domain representation. In sparse multipath scenarios, the channel output may be expressed as a sum of paths with individual delay and Doppler parameters. Our analysis shows that the channel response displays intrinsic differences when the delay is not an integer multiple of the sampling period and the Doppler shift is not an integer multiple of the frequency parameter $\Delta f=\frac{1}{LT_s}$. Compressed Sensing can enable significant improvements in sparse channel estimation. However, prior work on OTFS CS channel estimation has assumed that the delays are approximately integer multiples of the sampling period. In this paper, we developed a CS channel estimation scheme for OTFS systems that can model the \textit{continuous} delay and Doppler domain. Moreover, we have shown that in OTFS there are intrinsic differences when the delay coefficients are not integer, and the use of integer approximations results in catastrophic channel estimation errors. As a result, we propose an OMPBR algorithm that estimates delay and Doppler on a continuous domain. This significantly reduces the error in OTFS channel estimation with reasonable complexity. In future work we plan to extend our results to multiple-antenna joint sparse angular, delay and Doppler estimation problems and multi-user locations.

\bibliographystyle{IEEEtran}
\bibliography{{5GNR.bib,OTFS.bib,CompressiveSensing.bib}}

\end{document}